\documentclass[times,10pt,twocolumn]{article}
\usepackage{latex8}
\usepackage{times}
\usepackage{balance}

\usepackage{cite}
\usepackage{url}
\usepackage{booktabs}
\usepackage{nopageno}
\usepackage{fancyhdr}
\fancypagestyle{firstpagefooter}{
    \fancyhf{} % clear all header and footer fields
    \fancyfoot[L]{DOI: 10.18293/DMSVIVA2024-145} % put your DOI number here
}

\AtBeginDocument{%
  \providecommand\BibTeX{{%
    \normalfont B\kern-0.5em{\scshape i\kern-0.25em b}\kern-0.8em\TeX}}}

\usepackage{xspace}
\usepackage{caption}
\usepackage[skip=0cm,list=true,labelfont=bf]{subcaption}

\usepackage{soul}
\usepackage{lipsum,graphicx,multicol}
\usepackage{tabularx}

\usepackage{color, colortbl}
\definecolor{Gray}{gray}{0.9}
\usepackage[first=0,last=9]{lcg}

\usepackage[]{graphicx} % omit 'demo' for real document

\usepackage{algorithm}
\usepackage[noend]{algpseudocode}
\usepackage{nicefrac}
\providecommand{\keywords}[1]{\textbf{\textit{Index terms---}} #1}

\usepackage{amsmath}
\usepackage{algorithm}
\usepackage[noend]{algpseudocode}

\makeatletter
\def\BState{\State\hskip-\ALG@thistlm}
\makeatother

\begin{document}

%%
%% The "title" command has an optional parameter,
%% allowing the author to define a "short title" to be used in page headers.
\title{Automated Generation of Commit Messages in Software Repositories}
%%
%% The "author" command and its associated commands are used to define
%% the authors and their affiliations.
%% Of note is the shared affiliation of the first two authors, and the
%% "authornote" and "authornotemark" commands
%% used to denote shared contribution to the research.
\author{Varun Kumar Palakodeti, Abbas Heydarnoori\\
\emph{Department of Computer Science} \\
\emph{Bowling Green State University} \\
Bowling Green, USA\\
\{varunkp, aheydar\}@bgsu.edu\\
}

\maketitle

% Put the doi number on the lower left corner
% Apply the custom footer to the first page only
\thispagestyle{firstpagefooter}

%%
%% The abstract is a short summary of the work to be presented in the
%% article.
\begin{abstract}
  Commit messages are crucial for documenting software changes, aiding in program comprehension and maintenance. However, creating effective commit messages is often overlooked by developers due to time constraints and varying levels of documentation skills. Our research presents an automated approach to generate commit messages using Machine Learning (ML) and Natural Language Processing (NLP) by developing models that use techniques such as Logistic Regression with TF-IDF and Word2Vec, as well as more sophisticated methods like LSTM. We used the dataset of code changes and corresponding commit messages that was used by Liu et al.~\cite{NNGen}, which we used to train and evaluate ML/NLP models and was chosen because it is extensively used in previous research, also for comparability in our study. The objective was to explore which ML/NLP techniques generate the most effective, clear, and concise commit messages that accurately reflect the code changes. We split the dataset into training, validation, and testing sets and used these sets to evaluate the performance of each model using qualitative and quantitative evaluation methods. Our results reveal a spectrum of effectiveness among these models, with the highest BLEU score achieved being 16.82, showcasing the models' capability in automating a clear and concise commit message generation. Our paper offers insights into the comparative effectiveness of different machine learning models for automating commit message generation in software development, aiming to enhance the overall practice of code documentation. The source code is available at \cite{Repo}.
\end{abstract}

%%
%% Keywords. The author(s) should pick words that accurately describe
%% the work being presented. Separate the keywords with commas.

\section{Introduction}\label{sec:introduction}
Commit messages play a vital role in software development, documenting changes to facilitate an organized workflow and effective collaboration. They offer insights into the what, why, and how of code modifications, aiding in tracking progress, understanding the development process, and supporting debugging and future enhancements~\cite{Buse, Dyer, rostami2021ICSME}. The increasing complexity and scale of software projects underscore the necessity for commit messages, to ensure clear communication. Such documentation is crucial in open-source projects with diverse geographically contributors~\cite{Dyer}.

Crafting these messages manually in a fast-paced development environment and with varying documentation skills among developers leads to inconsistency, impacting code reviews, maintenance, and debugging~\cite{Dyer, Tian}. Moreover, under deadline pressures, developers might resort to generic descriptions, omitting crucial details~\cite{Dyer}. These challenges highlight the potential of Machine Learning (ML) and Natural Language Processing (NLP) to automate commit message generation, aiming to enhance both consistency and quality of software documentation~\cite{chen2022harnessing}.

Significant efforts have been made in ML/NLP for automating commit message generation, including information retrieval and LSTM methods. NMT~\cite{NMT} uses a neural machine translation algorithm to translate code diffs into commit messages. NNGen~\cite{NNGen} uses bag-of-words model and cosine similarity without a training phase. CoDiSum~\cite{CoDiSum} and RACE~\cite{shi2022race} introduce tree-based neural networks and retrieval-augmented methods to improve message accuracy and relevance.

Despite their innovation, existing ML/NLP models like NMT, RACE, and CoDiSum face challenges such as high computational demands data limitations, and scalability issues, especially with Nearest Neighbors algorithms~\cite{CoDiSum, NMT, shi2022race}.

We chose conventional machine learning models over modern transformer-based approaches to ensure efficiency and lightweight operation. The primary reason was the ability of these ML models to run on the M1 chip and other small computational devices. This approach allows for effective commit message generation without the need for high-performance GPUs, making it more accessible and practical for various environments. By focusing on ML models, we aimed to balance performance with resource efficiency, ensuring that our solution remains both powerful and adaptable to a wider range of hardware configurations~\cite{Mohsen2024DMSVIVA}.

Our approach trains ML/NLP models on a dataset of code changes and corresponding commit messages to learn patterns crucial for generating effective commit messages. Using the same dataset used by Liu et al.~\cite{NNGen}, we divided the data into training, testing, and validation phases. Our experiments, especially the Cosine Similarity with TF-IDF and Nearest Neighbors algorithm achieving a BLEU score of 16.82, showcase our model's ability to align closely with actual commit messages.

This paper explores lightweight ML/NLP approaches for commit message generation, evaluating their efficiency, effectiveness, and architectural nuances. Through comprehensive comparative analysis and evaluations, including BLEU scores and manual assessments, we provide insights into the models' performance and the potential of automating commit message generation.

This paper is organized as follows: Section~\ref{sec:proposed_solution} details our approach, including data preparation and model training. Section~\ref{sec:evaluations} evaluates the proposed approach. Section~\ref{sec:discussion} discusses the approach and results. Section~\ref{sec:threats} addresses threats to validity. Section~\ref{sec:related_work} reviews related work. Section~\ref{sec:conclusions} concludes the paper and provides future research directions.

\section{Proposed Solution}\label{sec:proposed_solution}

This section describes our approach, data preprocessing, data splitting, architecture, and training process of each model.

We focused on models like cosine similarity with TF-IDF, Word2Vec~\cite{Word2Vec} combined with Logistic Regression, and the PyTorch LSTM Model due to their efficiency and adaptability in processing high-dimensional spaces and leveraging pre-trained embeddings for enhanced performance. The effectiveness of these selected models and the challenges encountered with others such as XGBoost, SVM, Multinomial Na\"{\i}ve Bayes, and DistillBERT, which either fell short in scalability or were resource-intensive, are discussed comprehensively in subsequent sections.

\begin{table}[t]
\centering
\caption{Algorithms and Categories of Approaches}
\label{tab:algos}
\begin{tabular}{p{3.5cm}p{4.5cm}} % removed vertical lines and added @{} to remove padding at the start and end of the table
\toprule
\textbf{Category} & \textbf{Algorithm/Approach} \\
\midrule
Similarity-Based & K-Nearest Neighbors, Cosine Similarity with TF-IDF \\
Ensemble methods & XGBoost \\
Support Vector Machines & SVM \\
Probabilistic models & Multinomial na\"{\i}ve bayes \\
Neural Network models & Simple RNN with MPS, PyTorch LSTM Model, DistillBERT with Hugging Face Transformer \\
Vector Space models & Word2Vec with Logistic Regression, Logistic Regression with TF-IDF \\
Combined Similarity and Vector Space & KNN with TF-IDF \\
\bottomrule
\end{tabular}
\end{table}

Table~\ref{tab:algos} shows the categories of the approaches and highlights the algorithms used for each approach.

\subsection{Overview of the Proposed Solution}

Our solution integrates machine learning and natural language processing to craft commit messages, focusing on improved data preprocessing and model architecture. We aim to surpass the limitations of models like NMT, NNGen, RACE, and CoDiSum by enhancing accuracy, efficiency, and handling complexity, using standard evaluation metrics for performance comparison.

\subsection{Data Gathering}

We utilized the dataset referenced in Liu et al.~\cite{NNGen}, originally compiled by Jiang et al.~\cite{NMT}, a widely used benchmark in the field for evaluating new and existing methods.
%This dataset is accessible in our repository's data folder~\cite{Repo}.

\subsection{Data Preprocessing}
Our data preprocessing includes normalization steps such as whitespace stripping, lowercasing, tokenization, special character removal, stop word elimination, lemmatization/stemming, and vectorization. Except for LSTM, all machine learning models load data via a load\_data(path) function, with preprocessing handled by built-in Python functions and the NLTK library.

For LSTM, we wrote a custom script for preprocessing to convert text to indices and build a vocabulary. This involved tokenizing, lowercasing, removing special characters and stop words, and tagging the start and end of sentences. We then divided the data into training and test/validation sets.

A separate vocab.py script generated a vocabulary file, vocab.txt, ranking tokens by frequency and incorporating special tokens. Another script standardized input lengths to 100 tokens by padding or trimming, using the <PAD> token as filler. The processed, padded files serve as the final input for LSTM model training and testing.

\subsection{Model Architecture}

%\begin{figure*}[t]
%\centering
%\includegraphics[width=\textwidth]{pics/Architecture_of_Models.png}
%\caption{Architecture of ML Models}
%\label{fig:ml}
%\end{figure*}

Our models, leveraging Word2Vec embeddings and TF-IDF features, adopt straightforward architectures. The Cosine Similarity model, loads the data (training code changes and commit messages) and converts it into numerical vectors using TF-IDF.% as depicted in Fig.~\ref{fig:ml}.
The model, trained on the Cosine Similarity principle, identifies similar vectors and generates commit messages by selecting the nearest vector from the training data~\cite{Repo}.

In contrast, our Logistic Regression model with TF-IDF processes the data through a TF-IDF vectorizer and uses encoded commit messages for training. Implemented with PyTorch, it employs CrossEntropyLoss and Adam optimizer on Apple Silicon M1 GPU. This model outputs commit messages as text files. Both models are compared in Table~\ref{tab:TF-IDF}.

\begin{table*}[t]
\caption{Comparison of TF-IDF based models}
\label{tab:TF-IDF}
\begin{tabular}{p{0.2\linewidth}p{0.325\linewidth}p{0.4\linewidth}}

\toprule
\textbf{Feature} & \textbf{Logistic Regression with TF-IDF} & \textbf{Cosine Similarity with TF-IDF} \\
\midrule
Learning Type& Supervised Learning& Unsupervised Approach\\
Model Algorithm& Logistic Regression (PyTorch)& Cosine Similarity with Nearest Neighbors\\
Training Process& Trains on numerical labels of input training data& No training is required, training data for the similarity model\\
Hardware-utilized& Apple Silicon (mps)& Any CPU\\
Prediction Mechanism& Classifies instances into categories& Finds the nearest neighbor based on cosine similarity\\
Model Complexity& Involves weight adjustments and learning& Based on proximity in vector space\\
Interpretability& Can interpret feature importance& Straightforward (based on closest match in vector space)\\
\bottomrule
\end{tabular}
\end{table*}

The Logistic Regression model takes minutes to train and generate output, while the Cosine Similarity model takes seconds due to different learning techniques.

For Word2Vec, we use a distinct vectorization approach. Using Google's Word2Vec trained on Google News data, we employ a Cosine similarity-based Nearest Neighbors algorithm. The data undergoes vectorization to generate word embeddings, and the nearest vector match is used to generate commit messages. We also experimented with data preprocessing using the NLTK library for stop word removal, tokenization, and lower casing before vectorization.%, as illustrated in Fig.~\ref{fig:ml}.

To address scalability, we developed a Logistic Regression model integrated with Word2Vec, using similar preprocessing steps plus lemmatization. This model follows the same methodology as the TF-IDF model but uses Word2Vec embeddings.

\begin{table*}[t]
\caption{Comparison of Word2Vec based models}
\label{tab:word2vec}
\begin{tabular}{p{0.2\linewidth}p{0.325\linewidth}p{0.4\linewidth}}
\toprule
\textbf{Feature} & \textbf{Logistic Regression with TF-IDF} & \textbf{Cosine Similarity with TF-IDF} \\
\midrule
Learning Type& Supervised Learning& Unsupervised Approach\\
Model Algorithm& Logistic Regression (PyTorch)& Cosine Similarity with Nearest Neighbors\\
Training Process& Trains on numerical labels of input training data& No training is required, training data for the similarity model\\
Hardware utilized& Apple Silicon (mps)& Any CPU\\
Prediction Mechanism& Classifies instances into categories& Finds the nearest neighbor based on cosine similarity\\
Model Complexity& Involves weight adjustments and learning& Based on proximity in vector space\\
Interpretability& Can interpret feature importance& Straightforward (based on closest match in vector space)\\
\bottomrule
\end{tabular}
\end{table*}

Execution times vary, with Nearest Neighbors taking seconds and Logistic Regression taking minutes. Word2Vec offers extensive vocabulary and relies on a pre-trained model. A comparative analysis of Word2Vec-based models is given in Table~\ref{tab:word2vec}.

\begin{figure}[t]
  \centering
  \includegraphics[width=0.8\columnwidth]{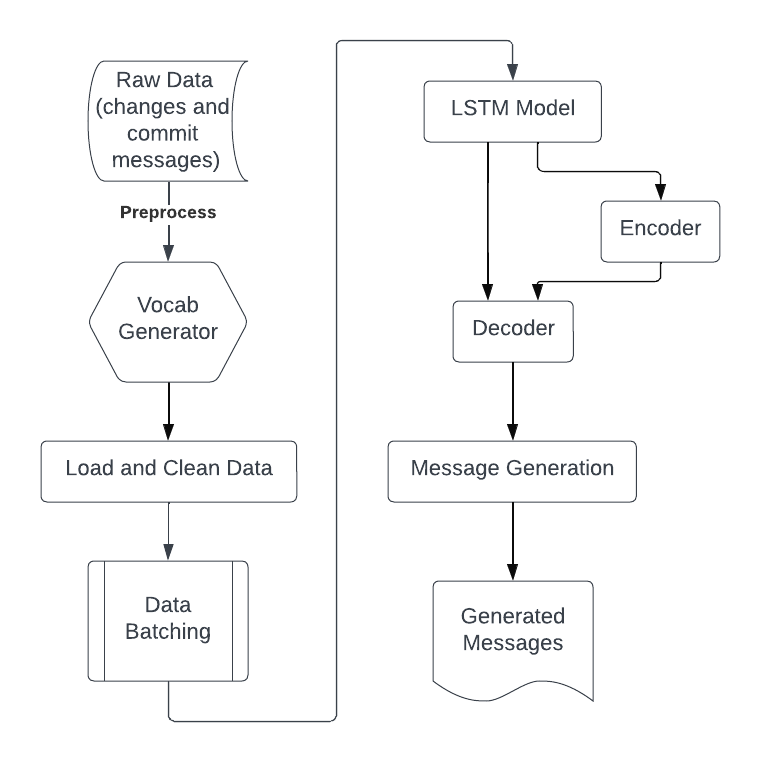}
  \caption{The architecture of the NLP model with the LSTM approach}
  \label{fig:lstm}
\end{figure}

Our LSTM-based model, built using PyTorch, is time-consuming and resource-intensive. Given the sequential nature of code changes, LSTM and Transformer models are ideal for capturing complexities in commit message generation. Preprocessing involves tokenization, stop word removal, and lemmatization. The data is then divided into training, testing, and validation sets, and standardized to a fixed length. The model’s architecture is shown in Fig.~\ref{fig:lstm}

The LSTM model’s Encoder processes the embeddings, creating a hidden state transferred to the Decoder, which generates commit messages sequentially. Training the LSTM model takes hours, with output generation taking minutes.

\section{Evaluations and Results}\label{sec:evaluations}

In this section we present an evaluation of the models, incorporating both qualitative and quantitative analyses to assess their effectiveness in generating commit messages.

\subsection{Research Questions}
Our study aims to answer the following research questions,

%\begin{itemize}
%\item
\textbf{RQ1: Which ML/NLP method is best for creating commit messages?} We will assess each model, comparing their BLEU scores and conducting manual evaluations.

%\item
\textbf{RQ2: How well do ML/NLP methods reflect human understanding of code changes?} Through manual reviews, we aim to gauge how these generated messages align with human perceptions and identify which outputs are preferred.

%\item
\textbf{RQ3: How do Large Language Models (LLMs) like ChatGPT compare with traditional ML/NLP methods?} We will prompt ChatGPT with code changes and compare its commit messages with those from our models in manual evaluations.

%\item
\textbf{RQ4: Can simpler, quicker methods compete with advanced models that require more resources?} We will test our streamlined models, designed for efficiency on the Apple M1 chip, against more complex, resource-intensive approaches.
%\end{itemize}

\subsection{Evaluations Setup}
Our experiments were conducted in a Python environment with all necessary libraries specified in the requirements.txt file of our repository~\cite{Repo}. We used a MacBook Air with an M1 chip for data preprocessing, model training, execution, and evaluation. PyTorch's MPS backend was used for leveraging the  M1 chip.

\subsection{Model Evaluation Criteria}
We used the BLEU score to evaluate the performance of machine-generated text by comparing n-gram overlap between model output and human output~\cite{BLEU02}. We calculated the BLEU score using the script from Liu et al.~\cite{NNGen}, initially developed by Jiang et al.~\cite{NMT}. We also conducted manual evaluations, categorizing diffs into small (under 50 tokens), medium (50 to 75 tokens), and large (over 75 tokens) sets for detailed analysis.

In all of the papers and approaches that we have come across, BLEU, Meteor and Rouge-L are the most common metrics used for evaluation~\cite{CoDiSum, shi2022race}. However, Rouge-L metric is usually chose for approaches involving summarization tasks~\cite{Rouge-L}, our approach is deals with machine translation and requires a task that can prioritize precision and emphasize the establishment of the relevancy of the machine generated text to the original text on which the machine is trained on. Meteor, the other common metric is a computationally complex approach, that deals with flexible matching where synonyms and stemming is taken into account when judging the precision of text generated or comparing the machine generated text to original~\cite{Meteor}. Hence, we use BLEU has higher accuracy when comparing the generated text to original as well as BLEU's approch penalizes when the words that are being compared are shorter than original~\cite{BLEU02}.

\subsection{Performance Results}

Each model had a unique way of producing the output, but most output messages generated had overlaps with other models, the highest being 16.82. Table~\ref{tab:bleu} can help understand the comparative scores of the models.

\begin{table}[t]
\centering
\caption{BLEU scores of models}
\label{tab:bleu}
\begin{tabular}{p{6.5cm}p{1.5cm}} % removed vertical lines and added @{} to remove padding at the start and end of the table
\toprule
\textbf{Model} & \textbf{BLEU Score} \\
\midrule
Cosine Similarity with TF-IDF and NN & 16.82 \\
Logistic Regression with TF-IDF & 16.13\\
Cosine Similarity with W2V and NN & 11.85\\
Cosine Similarity with W2V and NN (Preprocessed Data) & 15.01\\
Logistic Regression with W2V and Preprocessed Data & 3.17\\
LSTM Model& 0.68\\
\bottomrule
\end{tabular}
\end{table}

Model evaluation was based on their ability to generate commit messages for the test set of code changes. Each model's unique approach led to variations in performance as shown in the table.

For manual evaluation, a script selected random diffs, and we manually ranked the generated commit messages. Comments were provided for each diff which explains the choice.

\subsection{Quantitative Results}
The Cosine Similarity model with TF-IDF and Nearest Neighbors achieved the highest BLEU score of 16.82 and the Logistic Regression model with TF-IDF also performed well with a BLEU score of 16.13.

In contrast, models using Cosine Similarity with Word2Vec varied based on preprocessing, without preprocessing they scored 11.85, and with preprocessing improved to 15.01. The Logistic Regression model with preprocessed Word2Vec data scored 3.17. The LSTM model achieved a BLEU score of only 0.68, reflecting its limitations.

\subsection{Qualitative Results}

Our qualitative evaluation delved into the interpretation of commit messages generated by various machine learning algorithms, including LSTM and LLM technologies. Our qualitative evaluation categorized diffs into small, medium, and large sizes to assess model performance.

Small Diffs (Under 50 Tokens): The Logistic Regression model with TF-IDF consistently delivered outputs that closely matched the actual commit messages, often outperforming the more detailed LLM outputs.

Medium Diffs (50 to 75 Tokens): LLMs like ChatGPT generated commit messages that captured semantic depth, often surpassing other models, but Logistic Regression with Word2Vec model generated outputs that were aligned with actual messages.

Large Diffs (Over 75 Tokens): LLMs provided demonstrated contextual understanding of the code changes, although sometimes exceeded the desired brevity. They accurately identified specific components within diffs. All the other methods did not produce good commit messages.

\section{Related Work}\label{sec:related_work}

In this section, we analyzed the related work in the field of code documentation practice, focusing on and committing message generation for source code changes. We have outlined our understanding and analysis of the research work that was explored.

\subsection{Revisiting Learning-based Commit Message Generation}

Dong et al.~\cite{revisiting} underscored the importance of commit messages in software development, analyzing various methods for automated generation, particularly learning-based techniques. Their study examined the influence of datasets and model components on output quality, comparing rule-based, information retrieval-based, and learning-based approaches. They noted the advancements in deep learning have enhanced these techniques, though traditional models may obscure code semantics, leading to subpar performance. To address this, they introduced a two-stage generation paradigm that first creates an abstract representation of the commit message, with details added subsequently, thereby improving message relevance and clarity.

While Dong et al.~\cite{revisiting} proposed a two-stage generation process, our research implemented and evaluated different models that utilize TF-IDF, Word2Vec, and LSTM in the context of commit message generation. This not only demonstrates the practical application of their theoretical insights but also provides a comparative analysis of various models, contributing to a broader understanding of ML/NLP applications in software documentation. Our research stands out by offering empirical evidence on the effectiveness of these models and adding to the existing knowledge base in automated commit message generation.

\subsection{Model Architecture}

Jiang et al.~\cite{NMT} utilized a Neural Machine Translation (NMT) approach, leveraging an encoder-decoder architecture, to convert diffs into commit messages. They compiled a substantial, quality-controlled commit dataset from sizeable projects and trained their NMT model on this data~\cite{NMT}.

NNGen uses a bag-of-words model to convert diffs into vectors, then applies cosine similarity to find the most similar diffs using the Nearest Neighbors algorithm in the training set, streamlining the commit message generation process~\cite{NNGen}.

CoDiSum (Code Difference Summarization), a tree-based neural network model proposed by Shengbin Xu et al.~\cite{CoDiSum}, addresses limitations of prior commit message generation methods, particularly in understanding code structure and handling out-of-vocabulary (OOV) issues. CoDiSum’s encoder captures the syntactic framework and semantic representation of code, identifying and replacing code identifiers with placeholders. This process helps in generating more accurate commit messages. The decoder, a multi-layer unidirectional GRU, uses attention weights and a context vector to generate commit messages word by word, integrating a copying mechanism for OOV words. This mechanism calculates a distribution over the structure sequence, allowing the model to directly copy specific terms from the code, thereby maintaining technical accuracy. CoDiSum combines generation and copying probabilities, using a sigmoid function to determine the final word choice in the commit message.

The RACE (Retrieval-Augmented Commit Message Generation) model is a two-module approach for generating commit messages from code changes. The model utilizes Information retrieval of similar commits and a Neural Network approach for the generation of commit messages. The first phase is information retrieval where the model identifies the most similar code change from a large dataset using Cosine similarity~\cite{shi2022race}. The dataset consists of around a million pairs of code changes and corresponding commit messages. Once the information retrieval phase successfully finds a similar diff, it is passed to the second phase, where the encoder utilizes a Transformer-based architecture, Feed Forward Network. Based on the similarity of the input and the retrieved diff - commit message pair, the model decides how much the example should influence the new messages and produces an encoding, that will be passed to the decoder. The decoder, which is also a Transformer-based component, generates the commit message token by token. To generate the next token the model takes into consideration of the previous token to keep the commit message tokens meaningful~\cite{shi2022race}.

The models using the encode-decoder seq2seq approaches are complicated and require high-performance computational capacity compared to simple and lightweight approaches like ours. CoDiSum relies on understanding the semantics of the code changes, but our approaches are independent of the semantics so when our models encounter a new code change we base our output on the most similar vector whereas CoDiSum puts its efforts into understanding the code change's semantics if the semantics and logic of the code are complex the right commit message may not be generated. The major disadvantage of RACE is the reliance on information retrieval for commit message generation, our models in comparison are much more agile to any code change as they depend on vector similarity, and all of our models are straightforward and potentially faster to implement than RACE's two-module system.

\subsection{Data Preprocessing}

Data preprocessing is a critical phase in commit message generation, with each model adopting distinct but sometimes overlapping strategies. The NMT model focuses on cleaning commit messages by extracting relevant lines, removing identifiers, and avoiding large diffs, ensuring meaningful tokenization. CoDiSum goes further by extracting code structure and semantics, replacing identifiers with placeholders, and employing a copying mechanism for better message accuracy by including out-of-vocabulary words. NNGen simplifies diffs into vectorized ``bags of words,'' prioritizing term frequency over syntax or order. RACE's unique preprocessing uses token-level actions to depict code changes, utilizing tags to emphasize code modifications.

Compared to these approaches ours involves more thorough data cleaning processes, ensuring the removal of irrelevant or noisy data which is critical for the quality of the input. sophisticated tokenization strategies that go beyond simple white-space or punctuation-based methods. In the case of word2Vec, it inherently understands the syntactic and semantic aspects of programming languages. Our preprocessing steps are optimized to complement the architecture of your chosen ML models, enhancing their learning efficiency.

\section{Discussion}\label{sec:discussion}

Our models demonstrated unique performance in generating commit messages, with notable overlaps. The TF-IDF Logistic Regression model generally produced outputs closer to actual commit messages compared to others, including the Word2Vec Cosine Similarity model, where data preprocessing showed limited impact.

To answer our RQ1, regarding the ML/NLP method that is best for creating commit messages, the TF-IDF Cosine Similarity model led in BLEU scores, indicating strong performance in generating commit messages. However, manual evaluations involving the qualitative analysis revealed the TF-IDF Logistic Regression model produced more accurate commit messages for certain diffs as for the LSTM model, given the available data and computational resources, may not be the most effective approach.

Concerning RQ2, regarding how well do ML/NLP models reflect human understanding of the code changes, the interpretation of code changes varies significantly among individuals, often not aligning with actual or generated messages. Yet, for medium and small diffs, some model's outputs did match human interpretation. For example, the model using the TF-IDF with Logistic Regression has been noted to match the understanding of code changes in the same way that the authors of the original commit messages.

Regarding RQ3, Large Language Models (LLMs) like ChatGPT, outperformed all ML and NLP models in manual evaluations, demonstrating a human-like understanding unmatched by traditional approaches.

For RQ4, dealing with the conundrum of whether simple and quicker methods compete with advanced models that require high computational power, we want to emphasize that even though RACE model holds the highest BLEU score (25.66), our TF-IDF Cosine Similarity model surpassed other approaches like NMT, NNGen, and CoDiSum in BLEU score performance, achieving 16.82, yet RACE remains the top performer in this field. Hence, simple approaches that can be run locally cannot be dismissed when it comes to generation of commit messages.

Table~\ref{tab:all} shows the comparison of BLEU scores and compares the BLEU scores of the models that we designed to approaches that were contributed earlier to this domain. Table~\ref{tab:all} does not include the pre-trained models. The BLEU score for ChatGPT could not be included as it was not released.

Also, over the course of our research through the last year, there have been significant efforts in leveraging LLMs like ChatGPT for commit message generation. Researchers have explored various aspects of using ChatGPT, and reviewing their work, we have made some observations such as the context-aware, superior performance of ChatGPT. In 78\% of the evaluated samples, commit messages generated by ChatGPT were rated the best by human participants when compared to human-written commit messages, demonstrating their ability to produce high-quality, contextually accurate messages~\cite{chat}. Although LLMs are highly capable, their accessibility and security concerns are reasons to prefer alternatives where the models are deployed locally, like our approach. Considering a scenario where an enterprise environment that spans across the globe requires implementing a model for enhancing their code documentation practice, LLMs like ChatGPT would need access to proprietary information to generate commit messages. In this scenario, there would be a higher preference for a model that runs locally.

Even though a model can run locally, for example the Llama developed my Meta and the variants of Llama LLM were primarily developed to run locally, but its entire State Dictionary that is publicly available consists of billions of tokens~\cite{Llama}, making the model more generic to be an LLM, rather than our approaches that focus primarily on commit messages and code changes and are built on data gathered for this purpose.

A significant advantage of LLMs that was observed in our research and the research efforts of Zhang et al.~\cite{zhang} was that the LLMs outperformed all the traditional models when the code diff size is large, i.e. when the tokens in the code diff exceeds 100 tokens, as pointed out in our evaluation of the large category of diffs. Zhang et al.~\cite{zhang} also notes that large diffs are not a majority of the cases when it comes to code changes. Also, there is a significant gap in real time adaptation of LLMs for code documentation as noted by the researchers.

\begin{table}[t]
\centering
\caption{Comparison of BLEU scores of models}
\label{tab:all}
\begin{tabular}{p{6.5cm}p{1.5cm}} % removed vertical lines and added @{} to remove padding at the start and end of the table
\toprule
\textbf{Model} & \textbf{BLEU Score} \\
\midrule
RACE~\cite{shi2022race}& 25.66\\
Cosine Similarity with TF-IDF and NN& 16.82 \\
NNGen~\cite{NNGen}& 16.42\\
Logistic Regression with TF-IDF& 16.13\\
Lucene~\cite{Lucene}& 15.61\\
NMT~\cite{NMT}& 15.52\\
Cosine Similarity with Word2Vec and NN (Preprocessed Data)& 15.01\\
Cosine Similarity with Word2Vec and NN& 11.85\\
CommitGen~\cite{CommitGen}& 14.07\\
CoDiSum~\cite{CoDiSum}& 13.97\\
Logistic Regression with Word2Vec and Preprocessed Data& 3.17\\
LSTM Model& 0.68\\
\bottomrule
\end{tabular}
\end{table}

\section{Threats to Validity}\label{sec:threats}

In this section, we scrutinize the potential threats to the validity of our research, divided into internal validity, external validity, and reliability, to better understand their impact and strategize mitigations.

\subsection{Internal Validity}

One significant threat in our study is the integrity of the commit
message data. If the developer’s commit messages in the Java repos-
itories we collected do not accurately reflect the associated diffs,
our models could be trained on misleading data. The vast number of
commit messages makes it unfeasible to manually validate each one
for accuracy and relevance to its diff. Any discrepancies between
the commit messages and the actual code changes could lead to
models learning incorrect patterns, thus affecting the quality of the
generated commit messages.
But our source for the dataset is the same as the published work
of Liu et al.~\cite{NNGen} whose work, in turn, uses the dataset from Jiang
et al.~\cite{NMT}

\subsection{External Validity}

Our research is bound to the Java programming language, chosen
for its ubiquity and prevalence in software development~\cite{NMT}. This
choice means our trained models are fine-tuned to the patterns and
idioms of Java and may not generalize well to other programming
languages without additional modifications.
This language-specific focus presents a limitation in applying
our findings to the broader field of automated commit message
generation across diverse programming languages. To enhance
the external validity, future work should consider incorporating
datasets from various programming languages. This expansion
would necessitate adjusting preprocessing routines to accommodate
different syntactic and semantic structures, as well as retraining
the models to recognize and process language-specific constructs
accurately.

\subsection{Reliability}
Reliability concerns the reproducibility of our findings, manual evaluations can introduce subjectivity and bias. In our case, the evaluators are proven to show a significant level of experience and expertise with the Java programming language. Their review can be found in the repository~\cite{Repo}.

\section{Conclusions and Future Work}\label{sec:conclusions}

Generating commit messages poses a significant challenge,
necessitating messages that are not only comprehensible to
humans but also accurately reflect the context of code changes.
Our investigation into various machine learning algorithms
revealed that the Cosine Similarity with Nearest Neighbors
and Logistic Regression algorithms are notably effective and
computationally efficient for this purpose. However, their
performance excels predominantly with smaller diffs, aligning
closely with human interpretations of code changes. In contrast, Large Language Models (LLMs) like ChatGPT demonstrated superior performance for larger diffs. Our attempt to
develop a lightweight LSTM model, despite being tuned for
better performance, fell short of the effectiveness seen in
models utilizing high-performance GPUs. Interestingly, the model with the highest BLEU score did not always surpass
those with lower scores, especially within the realm of smaller
diffs, where the Logistic Regression with TF-IDF vectorizer
showed superior results.

Moving forward, we aim to broaden our exploration into
commit message generation, considering the potential of various pre-trained models to enhance our approach. The objective
is to develop a model that deeply understands the nuances
of diffs in a human-like manner, effectively utilizing new
and untrained words within the diffs. The ideal generated
message should be concise, not exceeding 72 characters, yet
meaningful. Additionally, optimizing the preprocessing of diffs
for model training and focusing on accurately capturing the
“what” aspect of changes remain priorities. This approach
seeks to improve the model’s preference among human evaluators, ensuring it is adaptable for training on both CPU and
GPU environments.

We shall explore modern technologies and LLMs like ChatGPT, Mistral AI and Meta-Llama and others. We have previously seen the potential of ChatGPT during out manual evaluation. We should explore models that have a potential for contextual awareness as to understand the semantic context of the code change that has been made, and generate an appropriate output in the form of commit message.

%%
%% The next two lines define the bibliography style to be used, and
%% the bibliography file.
\balance
\bibliographystyle{latex8}
\bibliography{main}

\end{document}